# Acoustic correlates of the syllabic rhythm of speech: Modulation spectrum or local features of the temporal envelope

Yuran Zhang[a], Jiajie Zou[a], Nai Ding[a,*]

[a] College of Biomedical Engineering and Instrument Science,

Zhejiang University, Hangzhou, Zhejiang, China

*Corresponding author:

E-mail address: ding_nai@zju.edu.cn (N. Ding).

College of Biomedical Engineering and Instrument Science,

Zhejiang University, Hangzhou, 310027, China

# Abstract

The syllable is a perceptually salient unit in speech. Since both the syllable and its acoustic correlate, i.e., the speech envelope, have a preferred range of rhythmicity between 4 and 8 Hz, it is hypothesized that theta-band neural oscillations play a major role in extracting syllables based on the envelope. A literature survey, however, reveals inconsistent evidence about the relationship between speech envelope and syllables, and the current study revisits this question by analyzing large speech corpora. It is shown that the center frequency of speech envelope, characterized by the modulation spectrum, reliably correlates with the rate of syllables only when the analysis is pooled over minutes of speech recordings. In contrast, in the time domain, a component of the speech envelope is reliably phase-locked to syllable onsets. Based on a speaker-independent model, the timing of syllable onsets explains about 24% variance of the speech envelope. These results indicate that local features in the speech envelope, instead of the modulation spectrum, are a more reliable acoustic correlate of syllables.

# 1. Introduction

Speech information is hierarchically organized into units of different sizes, e.g., phonemes, syllables, morphemes, words, and larger units such as phrases and sentences. The syllable is a mesoscale speech unit whose time scale lies between phonemes and morphemes, which are the most basic phonetic and semantic units, respectively. In the last 2 decades, the neural mechanisms underlying the neural representation of syllables have received a significant amount of attention: It is hypothesized that theta-band neural oscillations (4-8 Hz) provide a potential mechanism to parse a continuous speech stream into discrete syllable units (Assaneo and Poeppel, 2018; Ghitza, 2013; Giraud and Poeppel, 2012; Greenberg, 1999; Hovsepyan et al., 2020; Hyafil et al., 2015; Poeppel et al., 2008). According to this hypothesis, the syllable is a fundamental unit that transforms an auditory representation of speech into a linguistic representation. This hypothesis is mainly motivated by two observations. First, compared to phonemes and morphemes, syllables are perceptually more salient and can be sensed with relatively little linguistic knowledge (Liberman et al., 1974; Mehler et al., 1981). The likely reason why the syllables are perceptual saliency is that syllables are believed to have a reliable acoustic correlate, i.e., the speech envelope (Greenberg et al., 2003), which is defined as the low-frequency (typically below 16 Hz) fluctuations of sound power (Rosen, 1992), either within a narrow frequency band or across all frequencies (Fig. 1). The sound envelope is a basic acoustic feature that is well represented in the auditory cortex (Shamma, 2001), and is crucial for speech intelligibility (Chi et al., 2005, 1999; Elhilali et al., 2003; Elliott and Theunissen, 2009). Since it is often assumed that there is a simple relationship between speech envelope and syllable boundaries, it has been proposed that the brain may first determine the boundaries between syllables by analyzing the speech envelope, before further decoding the phonetic content based on more detailed analyses of spectro-temporal auditory features (Doelling et al., 2014; Ghitza, 2013).

Second, both the syllables and the speech envelope are quasi-rhythmic: The mean syllable rate generally falls between 4 and 8 Hz across languages (Coupé et al., 2019; Greenberg et al., 2003; Pellegrino et al., 2011), and the power of the speech modulation spectrum, i.e., the power spectrum of the speech envelope, also concentrates between 4-8 Hz across languages (Ding et al., 2017; Varnet et al., 2017). More importantly, when listening to speech, large-scale theta-band cortical activity measured by MEG and EEG is more strongly driven by the speech envelope (Ding and Simon, 2012; Lalor and Foxe, 2010) than phonemic and lexical features (Brodbeck et al., 2018; Daube et al., 2019). Such theta-band envelope-tracking response is also often interpreted as a neural marker for the syllable processing (Ding et al., 2016; Doelling et al., 2014) and is widely employed to probe the neural mechanisms underlying the speech encoding (Brodbeck and Simon, 2020; Ding and Simon, 2014) and to characterize the speech processing in special populations (Di Liberto et al., 2022; Palana et al., 2021).

Although the speech envelope is considered as a major acoustic correlate of the syllable rhythm, a close look at the literature reveals that this relationship is less well-established than typically assumed. For example, on the one hand, it is well-accepted that the rate of syllables matches the peak frequency of the speech modulation spectrum (Greenberg et al., 2003). On the other hand, it is suggested that the rate of syllables and peak modulation frequency have distinct properties: Some studies show that the mean rate of syllables greatly varies across languages (Coupé et al., 2019; Pellegrino et al., 2011), while other studies show that the peak modulation frequency is highly consistent across languages (Ding et al., 2017; Varnet et al., 2017). Second, the modulation spectrum of speech peaks around 4-5 Hz only when a 1/f trend is removed (Ding et al., 2017). Otherwise, the peak modulation frequency is around 0 Hz (Voss and Clarke, 1975). In contrast, no 1/f trend needs to be removed when calculating the mean rate of syllables. Therefore, it remains unclear whether the rate of syllables just coincidently matches the peak modulation frequency or whether the two signals truly correlate. The inconsistent results in the literature are further complicated by the fact that previous studies often analyze different speech corpora and employ different methods to calculate the rate of syllables and peak modulation frequency.

The rate of syllables, as well as the peak modulation frequency, is a summary statistic of a period of recording. Further evidence about the strong relationship between speech envelope and syllables comes from the instantaneous phase-locking between speech envelope and syllables: It is widely accepted that syllable boundaries correspond to the local minima in the speech envelope (Poeppel and Assaneo, 2020), and the syllable boundaries can be detected by analyzing the speech envelope (Hovsepyan et al., 2020; Hyafil et al., 2015; Villing et al., 2006). It is also obvious, however, that the speech envelope does not only encode information about syllable boundaries. For example, phonetic information is well encoded in the narrowband envelope (Stevens, 2002) and prosodic information, including duration information and gaps, can strongly influence both the narrowband and broadband envelope (Li and Yang, 2009). It remains unclear to what extent the syllable boundaries can explain the speech envelope, and whether the envelope features corresponding to syllable boundaries vary across languages and speaking styles.

Here, we revisit the relationship between the modulation spectrum and the rate of syllables by analyzing large speech corpora (>1000 hours of recordings). We first employ the large corpora to reevaluate whether the peak frequency of the grand-averaged modulation spectrum matches the mean rate of syllables across languages and speaking styles. Next, to test whether the peak modulation frequency is truly related to the rate of syllables, we quantify the correlation between peak modulation frequency and the rate of syllables across individual speakers. The peak modulation frequency and rate of syllables are summary statistics extracted from the speech envelope and syllable sequence, respectively. In the last step, we directly consider the time-domain speech envelope and syllable onset sequence, and quantify the phase-locking between the signals using computational models.

## 2. Frequency-domain relationship between speech modulations and syllables

### 2.1. Methods

#### 2.1.1. Speech corpus

Seven speech corpora were included in the analysis (Table 1), i.e., DARPA-TIMIT (Garofolo et al., 1993), Chinese-TIMIT (Yuan et al., 2017), Aishell-1 (Bu et al., 2017), WenetSpeech (Zhang et al., 2022), GigaSpeech (Chen et al., 2021), TED-LIUM (Rousseau et al., 2012), and Common-voice (Ardila et al., 2020). All corpora were transcribed. These corpora contained four speaking styles, i.e., read sentences, read audiobooks, public talks, and interviews (Table 1). All public talks were given by a single speaker, and interviews contained conversations between multiple speakers. For interviews, the recordings were manually selected from the YouTube category in GigaSpeech, with the criterion that no background music was present. For the corpus with multiple languages, i.e., Common-voice, we selected six languages for which the mean speaker duration was longer than 512 s, i.e., Thai, French, Spanish, Polish, Russian, and German. For corpora containing discourse-level recordings, we separated the recordings into sentences based on the sentence boundaries provided by the corpora. All sentences were within 20 s and resampled to 16 kHz. The scripts for corpus processing were available at https://github.com/austin-365/ms-tools.

### 2.1.2. Modulation spectrum

The speech envelope and modulation spectrum were extracted using the same methods in Ding et al. (2017), which were briefly outlined here (Fig. 1). The envelope was extracted by an auditory model that filtered the sound signal into 128 logarithmically distributed frequency bands between 180 Hz and 7 kHz. In each band, the filtered sound signal was half-wave rectified, smoothed, and decimated to 200 Hz. The output of the auditory model in each frequency band was referred to as the narrowband envelope, and the average of all 128 narrowband envelopes was referred to as the broadband envelope. The narrowband envelope characterizes sound power fluctuations within a narrow frequency band, while the broadband envelope better reflects coherent power fluctuations across frequency bands. The modulation spectrum was obtained by applying the Discrete Fourier Transform (DFT) to the square of the speech envelope. The narrowband spectrum was the root mean square (RMS) of the modulation spectra of the 128 narrowband envelopes, and the broadband spectrum was the modulation spectrum of the broadband envelope. The modulation spectrum was calculated for each sentence and the envelope of each sentence was adjusted to the same duration, i.e., 20 s, through zero-padding. The peak modulation frequency of a speaker was calculated using two methods, based on the modulation spectrum of individual sentences. Method 1 calculated the peak modulation frequency based on each sentence and then averaged the peak modulation frequency across sentences. Method 2 first calculated the RMS of the modulation spectra across all sentences and then extracted the peak of the RMS spectrum.

### 2.1.3. Rate of syllables

The syllable onsets were automatically extracted and the method was validated based on the corpora for which manual labeling was available (see Appendix). The boundaries between phonemes were extracted based on the audio and transcription using the Montreal Forced Aligner (MFA) (McAuliffe et al., 2017). The MFA also directly provided the syllable boundaries for Chinese, since syllables directly corresponded to the major unit in the writing system, i.e., characters. For English, the MFA provided the boundaries between words and phonemes, and the syllable boundaries were determined by grouping phonemes into syllables based on a dictionary, i.e., the Unisyn Lexicon (Fitt, 2001). For other languages, the syllable boundary was not extracted, attributable to the diversity in syllable structure.

The rate of syllables was characterized using three measures. The first measure was the syllable rate, i.e., the number of syllables divided by the duration of the recording, and the second measure was the articulation rate, i.e., the number of syllables divided by the total duration of syllables. The difference between the syllable rate and articulation rate was that the former measure was affected by the silence periods in speech but the latter measure was not. The third measure was syllable mode, i.e., the mode of the histogram of the reciprocal of the duration of individual syllables (Greenberg et al., 2003), which was estimated by the Gaussian kernel density method with the window width determined using Scott's rule (Scott, 2015).

## 2.2. Results

We first compared the shape of the speech modulation spectrum and the distribution of syllable duration. In this analysis, all measures were averaged within each corpus. On average, the broadband and narrowband modulation spectrum peaked at 3.9 Hz (SD = 0.7 Hz) and 4.8 Hz (SD = 0.7 Hz), respectively. The mode of syllable rate distribution, as well as the mean syllable rate and mean articulation rate, coincided with the peak frequency of the modulation spectrum (Fig. 2A). The mean difference between the different measures of the rate of syllables and peak modulation frequency was shown in Fig. 3A, which was generally within 1 Hz (M = 0.5 Hz, SD = 0.3 Hz). The smallest difference was observed between the articulation rate and the peak frequency of the narrowband modulation spectrum calculated using Method 2.

Next, we asked whether the rate of syllables and peak modulation frequency were correlated across individual speakers. The correlation coefficients between different measures were averaged across corpora and shown in Fig. 3B. The highest correlation was observed between the articulation rate and the peak frequency of the broadband modulation spectrum calculated using Method 1, and this correlation was further illustrated for individual corpora in Fig. 2B. The correlation was the lowest for the two smaller corpora, i.e., TIMIT and Chinese-TIMIT (< 6 hours of recording), and the highest for AISHELL (179 hours of recording). It is worth noting that both Chinese-TIMIT and AISHELL contained recordings of read Chinese sentences, and mainly differed in the recording duration per speaker. To test whether the recording duration constrained the correlation between peak modulation frequency and articulation rate, we calculated the correlation coefficient based on subsets of the recordings. It was observed that the correlation increased when the recording duration of individual speakers increased, and this result is consistent for Chinese and English (Fig. 4), as well as six other languages (Fig. S5). When the recording duration per speaker was 4 s, the correlation between peak modulation frequency and articulation rate was low ($R = 0.25$ on average, SD = 0.05 across corpora), while the asymptote correlation was above 0.5 for most corpora ($R = 0.61$ on average, SD = 0.11 across corpora). Based on a sigmoid fit, the correlation coefficient reached 95% of the maximum value when the recording duration was beyond 512 s for most corpora.

## 3. Time-domain relationship between speech modulations and syllables

The correlation between peak modulation frequency and the rate of syllables characterized whether the envelope rhythm and the syllable rhythm had the same center frequency. Having the same frequency range, however, did not guarantee that the two sequences were time-locked. Furthermore, even when two signals did not have the same center frequency, they might still have components that were phase-locked. Therefore, in the following, we directly analyzed the phase-locking between speech envelope and syllable onsets using two predictive models, i.e., the Temporal Response Function (TRF) model and Long Short-Term Memory (LSTM) network (Fig. 5A).

### 3.1. Methods

#### 3.1.1. Temporal response function

The phase-locking between speech envelope and syllable onsets was characterized using the TRF and LSTM models. The TRF described the relationship between two sequences using a linear time-invariant system model (Crosse et al., 2016; Ding and Simon, 2012). In brief, the TRF is an optimal linear filter to transform one signal into another. The syllable onsets were encoded by a binary sequence whose value was 1 only at the onset of a syllable (sampling rate 50 Hz), and the broadband envelope was also decimated to 50 Hz to facilitate training. If the syllable onset sequence and the broadband envelope were referred to as $\sigma(n)$ and $env(n)$ respectively, their relationship was characterized using the following equation:

$$env(n) = \sum_m TRF(m)\sigma(n-m) + \varepsilon(n),$$

where $TRF(m)$ and $\varepsilon(n)$ referred to the TRF and the residual error of the model, respectively. The range of time lags considered in the model, i.e., $m$, was between -0.5 s and 0.5 s. The TRF was estimated using ridge regression with 10-fold cross-validation. The predictive power of the TRF was defined as the correlation coefficient between the predicted envelope and the actual broadband envelope, which was averaged over the 10 folds. We considered a speaker-independent TRF model and trained the TRF based on each corpus, with all the sentences in the training set being concatenated. Pilot analyses showed that the speaker-independent TRF could predict the speech envelope with similar accuracy as a speaker-dependent TRF trained based on each speaker (data not shown).

### 3.1.2. Long Short-Term Memory

The LSTM (Hochreiter and Schmidhuber, 1997) was a deep neural network model transforming an input sequence into an output sequence (see Fig. S1). The TRF was a time-invariant model: Each syllable onset triggered the same speech envelope. The LSTM model, however, was context-dependent: The speech envelope triggered by a syllable onset depended on the context, i.e., onsets of neighboring syllables, and the range of the context was automatically determined within the model. We employed an 8-layer bidirectional LSTM to transform the syllable onset sequence into the speech envelope. A speaker-independent LSTM model was learned based on each corpus. Since it was time-consuming to train the LSTM model, we broke each corpus into a training set, a validation set, and a test set. The training set contained 80% of the sentences randomly chosen from the corpus, while the validation set and the test set each contained half of the remaining 20% of sentences.

### 3.1.3 Statistical analysis

To evaluate whether syllable onsets could significantly predict the speech envelope by either the TRF or the LSTM model, we estimated the chance-level predictive power with a permutation strategy. Specifically, we randomly paired the syllable onset sequence of a sentence with the speech envelope of another sentence, and calculated the prediction based on the TRF or LSTM model. This procedure was repeated 1000 times, creating 1000 chance-level predictive power. If the actual predictive power was lower than $A$ out of 1000 chance-level prediction performance, the significance level was $(A+1)/(1000+1)$.

### 3.2. Results

The TRF showed a trough at $32 \pm 8$ ms latency after a syllable onset (Fig. 5B), indicating a local minimum in the broadband envelope after a syllable onset. The predictive power of TRF, i.e., the correlation coefficient between the actual envelope and the envelope predicted by syllable onsets, generally ranged between 0.25 and 0.35 for each corpus (Table 2, $p < 10^{-3}$ for each corpus, FDR corrected). The square of the predictive power, i.e., $R^2$, for each speaker was $10.6 \pm 4.4\%$, which could be interpreted as the percent of envelope variance explained by the syllable onset. For the context-dependent LSTM model, the predictive power, i.e., $R$, generally fell between 0.4 and 0.6 for each corpus (Table 2, $p < 10^{-3}$ for each corpus, FDR corrected), about 1.5 times higher than the predictive power of the TRF. The square of the predictive power, i.e., $R^2$, for each speaker was 24.2 ± 6.9%. Furthermore, the models, i.e., TRF and LSTM, learned from the two smaller corpora could generalize well to the larger corpora, even across languages (Table 2, $p < 10^{-3}$ for each corpus, FDR corrected). Finally, to analyze whether a long recording duration was required to reveal the phase-locking between speech envelope and syllable onsets, we calculated the predictive power of both models on subsets of large corpora. It was observed that the predictive power did not clearly change with the recording duration (see Fig. S2). These results suggested that the phase-locking between speech envelope and syllable onsets was robust to recording duration and both models could generalize to the speech of different styles and languages.

## 4. Discussion

The current study investigates the relationship between speech envelope and syllable boundaries. In contrast to the common belief that the peak modulation frequency is an acoustic correlate of the rate of syllables (Greenberg et al., 2003), the study shows that the peak modulation frequency is not strongly correlated with the rate of syllables for individual speakers, especially when only a few seconds of recordings are analyzed per speaker. Therefore, the peak modulation frequency cannot serve as a useful cue to determine the rate of syllables of individual sentences. In contrast, the TRF/LSTM analyses reveal that, in the time domain, the broadband speech envelope is reliably phase-locked to the onset of individual syllables. Therefore, to encode the syllable rhythm, the brain may have to extract the syllable-relevant temporal features from the speech envelope, instead of just passively following the speech envelope.

### 4.1. Speech modulation spectrum

The narrowband modulation spectrum has been widely applied to characterize the acoustic rhythm of speech, and it has a similar shape across languages and common speaking styles, e.g., book reading for adults and interviews (Ding et al., 2017; Varnet et al., 2017). More specialized speaking styles including infant-directed speech (IDS), however, can modulate the shape of the modulation spectrum. In the following, we first summarize previous studies on common speaking styles and then discuss findings on more specialized speaking styles (Fig. 6). The mean peak frequency in the modulation spectrum is 4.5 Hz for the English telephone corpus Switchboard (Greenberg et al., 2003), between 4.4 and 4.8 Hz for English corpora with different speaking styles (Ding et al., 2017), between 4.3 and 5.4 Hz for naturalistic discourse-level speech across nine languages (Ding et al., 2017), and between 4.3 and 5.5 Hz for semi-read speech (SRS) corpus in ten languages (Varnet et al., 2017). In the current study, the peak frequency of the narrowband modulation spectrum is between 4.2 and 5.9 Hz across corpora, with an average of 4.9 Hz (Table 3). The broadband modulation spectrum also shows a peak, and the peak frequency ranges from 3.5 to 4.5 Hz across six languages (Poeppel and Assaneo, 2020). In the current study, the peak frequency of the broadband modulation spectrum is between 3.3 and 5.0 Hz across corpora, with an average of 4.3 Hz (Table 3). The peak frequency of the broadband modulation spectrum is lower than that of the narrowband modulation spectrum in both the current study and previous studies (Poeppel and Assaneo, 2020). Therefore, for both the broadband and narrowband modulation spectrum, the peak frequency is highly consistent across languages and speaking styles and the variation across studies is within 1.7 Hz.

When the speaker attempts to speak in a particularly clear way, the modulation spectrum of speech is altered. In particular, low-frequency temporal modulations below 4 Hz are enhanced across such speaking scenarios, including speaking to an infant (Leong et al., 2017), a child (Leong and Goswami, 2015; Pérez-Navarro et al., 2022), or a hearing impaired person (Krause and Braida, 2004), as well as speaking in a noisy environments (Bosker and Cooke, 2018). In other words, when the listener is expected to experience difficulties in understanding speech, due to either degraded hearing conditions or the lack of language knowledge, the speaker tends to modify their speaking styles to enhance low-frequency modulations, which can be achieved even when the syllable rate is minimally changed (Bosker and Cooke, 2018; Krause and Braida, 2004; Pérez-Navarro et al., 2022). The most exaggerated changes in both low-frequency modulations and the syllable rate occur for infant-directed speech (Fig. 6), when the listener is expected to have the lowest ability to understand speech.

### 4.2. Rate of syllables

We also summarized the rate of syllables reported in the literature in Fig. 6. The mean syllable rate is between 5.2 and 7.8 Hz across eight languages in one study (Pellegrino et al., 2011) and is between 4.7 and 8.0 Hz across seventeen languages in another study (Coupé et al., 2019). In the current study, the syllable rate is between 3.5 and 6.1 Hz across eight languages (Table 3). Similarly, based on the literature, the articulation rate greatly varies across speaking styles. It ranges from 3.1 to 5.6 Hz for different speaking styles in English (Jacewicz et al., 2009) and from 3.8 to 7.2 Hz in German (Jessen, 2007). In the current study, the articulation rate ranges from 4.3 to 5.2 Hz across speaking styles for English and from 4.3 to 6.0 Hz for Chinese (Table 3). Finally, the syllable mode is only previously reported for Switchboard, i.e., 5.2 Hz

(Greenberg et al., 2003), and varies between 3.5 and 5.3 Hz across corpora in the current study. In summary, across different languages and speaking styles, the reported syllable rate varies from 3.5 to 8.0 Hz, and the reported articulation rate varies from 3.1 to 7.2 Hz.

The results reported in the literature are based on different corpora, which makes it challenging to compare the results across studies. For example, the articulation rate is by definition faster than the syllable rate, while the articulation rate reported in the literature is often slower than the syllable rate. A likely reason is that the studies reporting high syllable rates are studies that ask the speaker to read after getting familiar with the materials (Coupé et al., 2019). The comparison between languages is especially challenging given the potential influences of speaking styles. For example, one study reveals a faster articulation rate in Chinese than English (Ann Burchfield and Bradlow, 2014), while other studies report the opposite (Coupé et al., 2019; Pellegrino et al., 2011). The current study analyzed two corpora for read Chinese sentences and one corpus for read English sentences, and the articulation rate of the English corpus falls between the articulation rates of the two Chinese corpora. Therefore, although different languages are reported to have distinctions in their mean syllable rate, such distinctions have to be validated based on larger-scale controlled studies.

### 4.3. Relationship between modulation spectrum and rate of syllables

Although it is commonly assumed that the modulation spectrum reflects the distribution of the syllable rate (Greenberg et al., 2003), here it is demonstrated that the peak frequency of the narrowband modulation spectrum only weakly correlates with the mode of the syllable rate distribution ($R$ around 0.3, Fig. 3B). Therefore, the peak frequency of the narrowband modulation spectrum is not strongly influenced by the rate of syllables, but instead may originate from the biophysical properties of the human articulator (Chandrasekaran et al., 2009). The peak frequency of the broadband modulation spectrum correlates better with the rate of syllables of each speaker (Fig. 3B), especially when minutes of recordings are analyzed per speaker. The results suggest that the broadband envelope, which reflects intensity fluctuations that are coherent across frequency bands (Poeppel and Assaneo, 2020), is a better indicator of the syllable rhythm, compared with the envelope within each narrow frequency band. The broadband modulation spectrum, however, also weakly correlates with the rate of syllables when only a few seconds of speech recordings are analyzed.

When a speech signal is uniformly time-compressed to increase the rate of syllables, the peak modulation frequency faithfully tracks the rate of syllables (Fig. 7A, left). In other words, if fast speech is simply a time-compressed version of slower speech, there will be a perfect correlation between the rate of syllables and peak modulation frequency ($R = 1$). Then, why is the correlation so low for real sentences? One possibility is that the low-frequency temporal modulations in speech are in fact stronger than the modulations near the rate of syllables (Voss and Clarke, 1975). The strong low-frequency power may render the syllable-rate peak in the modulation spectrum highly variable across utterances, even after the 1/f trend is corrected. A second reason is that when the speech rate changes, different speech segments do not uniformly scale: For example, previous studies have also widely documented that an increase in the speech rate is accompanied by, e.g., vowel reduction (Lindblom, 1963; Taylor et al., 2014), consonant reduction (van Son and Pols, 1999), an increase in co-articulation (Agwuele et al., 2008), and lightly changes in the consonant/vowel ratio (Kessinger and Blumstein, 1998).

Finally, another possibility is that the phonetic content of speech matters, since in the corpora analyzed in the current study the phonetic content of speech varies across speakers. To test whether the phonetic content can indeed explain why the peak modulation frequency is not reliably correlated with the articulation rate for individual sentences, we further analyze the two sentences that were spoken by all speakers in the TIMIT corpus. For these two sentences, however, the articulation rate is still poorly correlated with the peak modulation frequency across speakers (left panels of Fig. 7B, $R = 0.28$ and $0.11$ for the two sentences, respectively). This result suggests that, even when the phonetic content of speech is kept constant, the articulation rate of a single sentence is still not strongly correlated with the peak modulation spectrum. Analyses of the two sentences, however, provide additional insights into what the modulation spectrum reflects and why the peak modulation frequency is not reliably correlated with the articulation rate for short utterances. In the middle panels of Fig. 7, we show the modulation spectrum of each speaker, and the speakers are sorted based on the articulation rate. The modulation spectrum averaged across fast, typical, and slow speakers is also shown (Fig. 7, right panel). These results demonstrate that the modulation spectrum has different shapes for the two sentences, and the difference is reliable across speakers. The modulation spectrum of a sentence is distinguishable from the modulation spectrum averaged over a corpus, and tends to have more narrow peaks than the averaged spectrum, which are signatures related to the phonetic content. Furthermore, for each sentence, the modulation spectrum of a speaker is clearly influenced by the articulation rate. Essentially, when the articulation rate increases, all peaks in the modulation spectrum shift to higher modulation frequencies (middle column in Fig. 7). One reason why the peak modulation frequency is not highly correlated with the articulation rate is that the modulation spectrum of a sentence has multiple peaks and the strongest peak varies across speakers. Consequently, although each peak in the modulation spectrum tracks the articulation rate to some extent, the frequency associated with the highest peak greatly varies across speakers. In summary, since the phonetic content strongly affects the modulation spectrum of the sentence and results in multiple spectral peaks, the frequency of the highest peak poorly correlates with the articulation rate. When the modulation spectrum is averaged over a large number of utterances, either from the same speaker or across speakers, it has a stereotypical bell shape and the frequency of this single spectral peak reliably correlates with the articulation rate.

**4.4. Relationship between syllable onsets and speech envelope**

The TRF analysis reveals that, even with a short 4-s speech recording, the broadband envelope of speech shows reliable phase-locking to the onset of individual syllables (see Fig. S2). The TRF shows that the envelope tends to show a local minimum at about 32 ms after a syllable onset and a local maximum about 160 ms after a syllable onset. The results are consistent with previous findings that the syllable boundaries can be extracted from the speech envelope (Hovsepyan et al., 2020; Hyafil et al., 2015; Villing et al., 2006), and that a syllable typically has one local maximum in the envelope, which corresponds to the nucleus (Greenberg et al., 2003; Hooper, 1976; Oganian and Chang, 2019), while the boundaries of syllables more closely correspond to troughs in the envelope (Mermelstein, 1975). The current study, however, extends the previous studies by quantifying how much percent of the variance of the speech envelope can be explained by syllable onsets. Furthermore, previous studies that employ the speech envelope to extract syllable boundaries often view the results as correct if the extracted boundary is within 50 ms (Hovsepyan et al., 2020; Hyafil et al., 2015; Räsänen et al., 2015; Villing et al., 2006) from the human-annotated boundaries, while the current study quantifies the timing between the troughs in envelopes and the syllable onsets.

The TRF model is a time-invariant model, which predicts that each syllable onset triggers the same speech envelope, e.g., the same duration, amplitude, and time course, regardless of the phonetic content and the context. Although this assumption is clearly oversimplistic, based on the TRF model, the onset of individual syllables explains about 11% of the variance of the broadband speech envelope, i.e., $R^2$. Furthermore, when considering contextual information, i.e., the speech envelope triggered by a syllable onset is influenced by the timing of other syllables, syllable onsets can explain about 24% of the variance of the speech envelope, i.e., $R^2$, by the LSTM model. This ratio is high in the sense that the model does not consider any phonetic information. Furthermore, both the TRF and LSTM models generalize well between English and Chinese and across speech corpora of different speaking styles, suggesting that the time-domain phase-locking between speech envelope and syllable onset is potentially a universal feature.

### 4.5. Implications for the acoustic cues for syllable boundaries

The speech envelope, as well as the modulation spectrum, is an acoustic feature of speech, while the syllable is a linguistically defined unit. How to link the acoustic features of speech to linguistic units is one of the most central questions in the psycholinguistics and neurolinguistics (Poeppel et al., 2008; Poeppel and Embick, 2017). The difficulty to identify reliable acoustic features of phonemes (Liberman et al., 1967), as well as the lack of awareness of phonemes in illiterate people (Morais et al., 1979), motivates studies to investigate whether other linguistic units have more reliably acoustic correlates. Since the speech envelope, as well as the modulation spectrum, has been suggested to provide straightforward acoustic correlates for syllables (Greenberg et al., 2003), the syllable has been considered as an effective interface to link auditory and language processing (Ghitza, 2013; Giraud and Poeppel, 2012).

The current study confirms that the speech envelope is phase-locked to the syllable onsets (Poeppel and Assaneo, 2020), but also shows that only a component in the speech envelope is tracking the syllable onset: Based on the LSTM model, syllable onsets only explain about a quarter of the variance of the envelope ($R^2$). Similarly, the modulation spectrum is sensitive to a number of factors, including phonetic and prosodic features, and only reliably tracks the rate of syllables when averaged over long utterances. These results suggest that the brain still needs to apply additional computations to extract the syllable boundaries based on the speech envelope. A related point is that when the listener is expected to be less efficient at understanding speech, the speakers tend to modify their speaking style (see section 4.1). It is possible that speech produced in such conditions, including infant/child-directed speech, has a more transparent link between the speech envelope and syllable boundaries, so that the listener can more easily map the acoustic features of speech onto linguistic units (Goswami, 2019). Furthermore, it has recently been suggested that a number of clinical populations show deficits in the cortical encoding of the speech envelope (see Palana et al., 2021 for a review), and such populations include dyslexia children/adults and sub-clinical poor readers (Di Liberto et al., 2018; Lizarazu et al., 2021; Molinaro et al., 2016). In these populations, it is possible that degraded neural encoding of the speech envelope renders it even more challenging to extract linguistic units such as syllables. Future studies are needed to test whether these populations can benefit from speech that has a more transparent mapping between envelope and syllables.

Lastly, although the current study focuses on the relationship between speech envelope and syllables, the envelope, especially the temporal modulations below 4 Hz, also provides useful cues for other levels of linguistic units, such as stressed syllables (Greenberg et al., 2003) and intonational phrases (Inbar et al., 2020). For child-directed speech, in particular, it has been observed that the very low-frequency modulations around 2 Hz provide cues for prosodic stress while modulations around 4 Hz provide cues about the syllables (Leong and Goswami, 2015). Therefore, the hierarchy of temporal modulations may correspond to the hierarchy of speech prosody, and such mapping between acoustic and linguistic information is hypothesized to play a critical role in facilitating language acquisition (Goswami, 2019). Future studies are needed to quantify the relationship between speech envelope and the full hierarchy of prosodic units, and whether the relationship differs between infant-, child-, and adult-directed speech.

In summary, a component in the speech envelope is phase-locked to the syllable onsets. This component, however, is not always the most powerful in speech, and consequently, the peak modulation frequency only poorly correlates with the rate of syllables for single sentences that typically last for a few seconds. When the modulation spectrum is averaged over many minutes of speech recordings, however, the signatures related to the phonetic content of single sentences smear out and the peak modulation frequency can strongly reflect the mean rate of syllables.

## Acknowledgements

This work was partly supported by the National Natural Science Foundation of China (32222035) and Key Research and Development Projects of Zhejiang (no. 2022C03011).

**Figure captions**

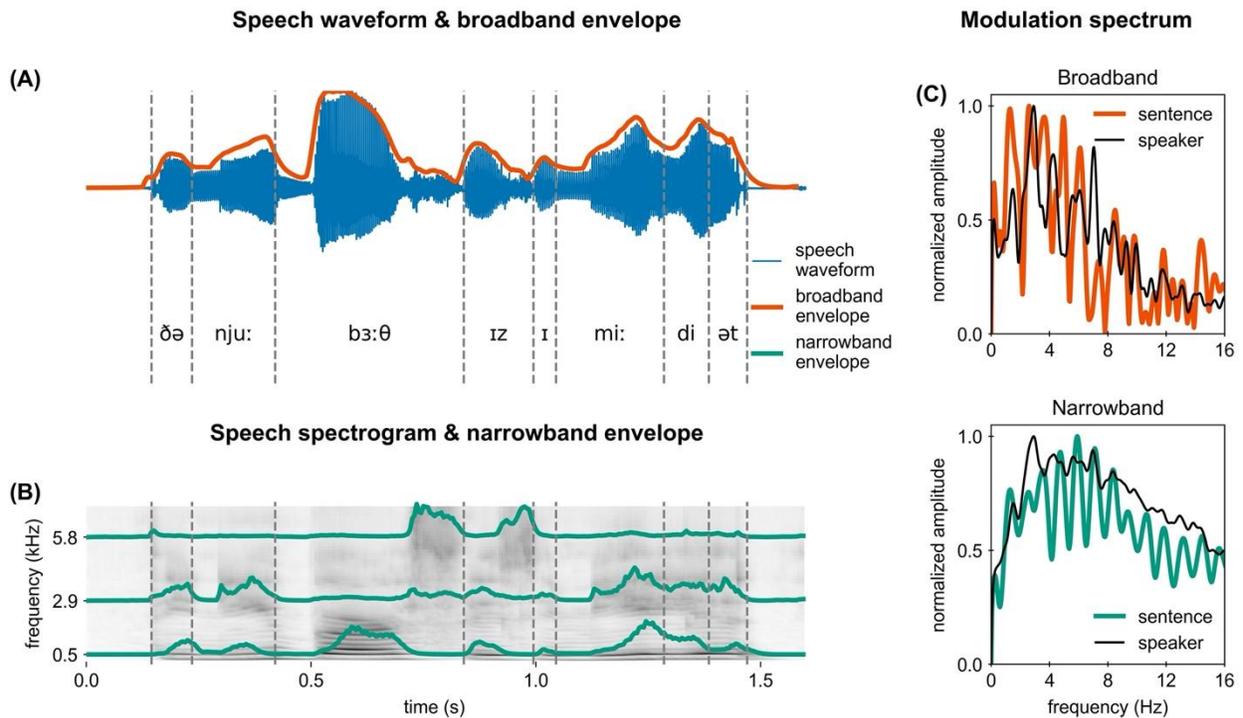

**Figure 1.** Speech envelope and modulation spectrum. (A) The speech waveform and broadband envelope of a sentence ("the new birth is immediate") from a speaker. Syllable boundaries are marked by dashed lines. (B) Speech spectrogram and narrowband envelope. Each row in the spectrogram constitutes a narrowband envelope. (C) Normalized modulation spectrum. The broadband and narrowband modulation spectra are calculated based on the broadband and narrowband envelopes, respectively. The colored and black lines are calculated based on that sentence and all sentences of that speaker, respectively.

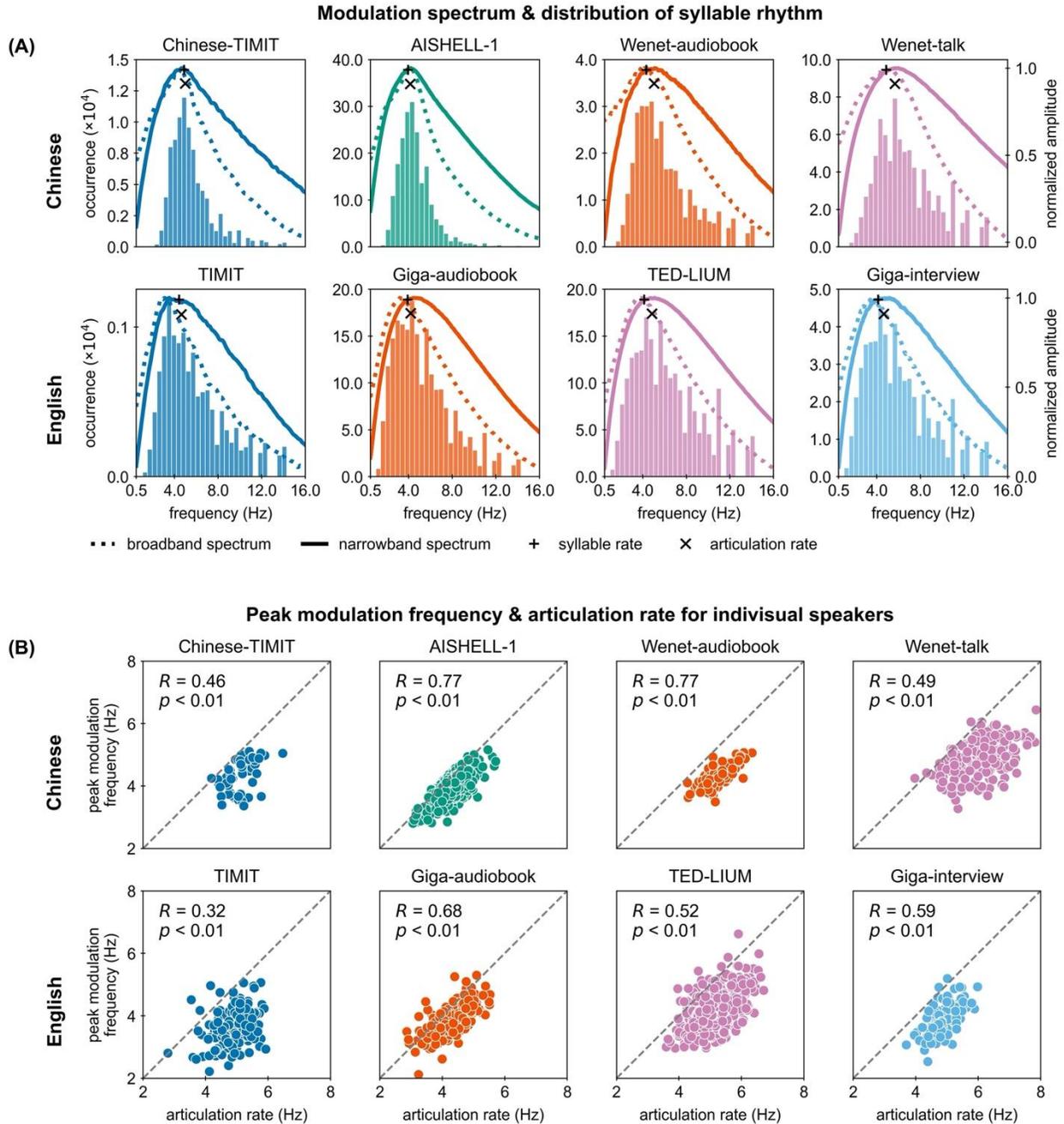

**Figure 2.** Peak modulation frequency, rate of syllables, and their correlation. (A) The broadband and narrowband modulation spectra, as well as the histogram of the reciprocal of syllable durations, are shown for each speech corpora. The mean syllable rate and the mean articulation rate are shown by the black markers. The syllable mode is the peak of the histogram. For each language, each corpus is shown with a unique color which is consistent in all figures. (B)

Correlation (*R*) between peak modulation frequency and articulation rate for individual speakers. The peak modulation frequency is extracted from the broadband modulation spectrum using Method 1. Each dot denotes an individual speaker.

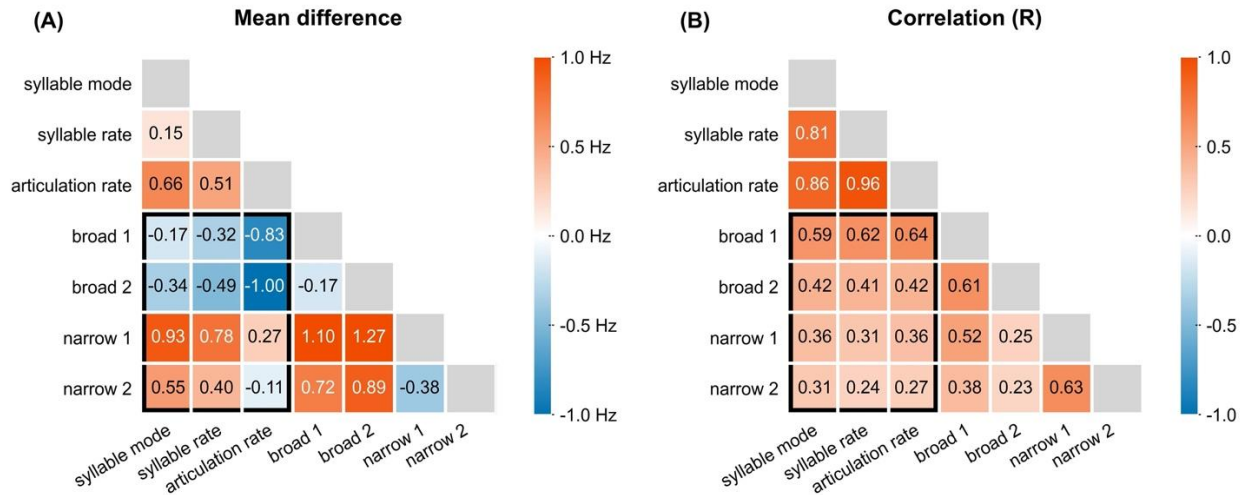

**Figure 3.** Mean difference and correlation between different measures describing the rate of syllables and peak modulation frequency. (A) Each measure is averaged across corpora and the difference between measures is shown. (B) Pairwise correlation (*R*) of different measures is calculated within each corpus, and the correlation coefficient of each corpus is weighted by the total duration of the corpus and averaged. The black box highlights the relationship between measures of the rate of syllables and measures of the peak modulation frequency.

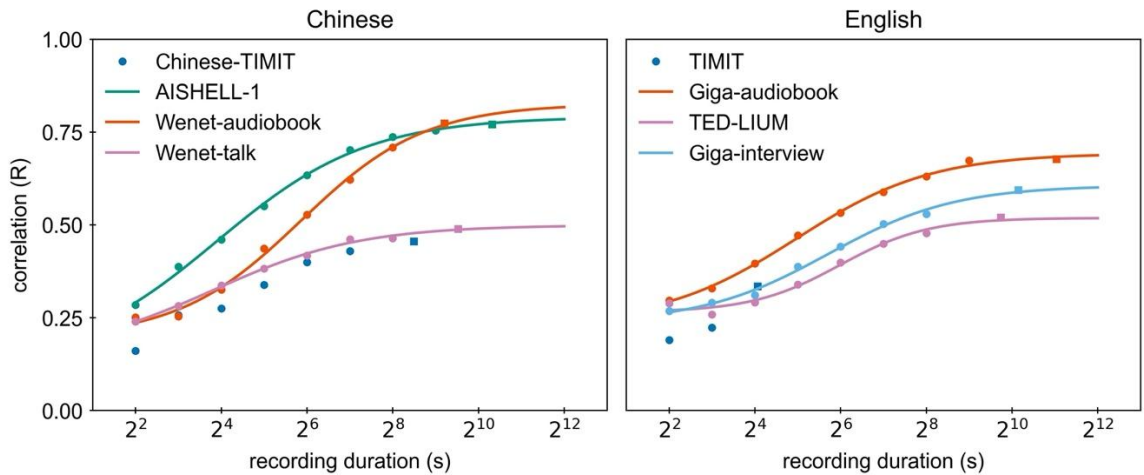

**Figure 4.** Correlation (*R*) between peak modulation frequency in broadband and articulation rate as a function of recording duration. The recording duration is the duration of the speech recordings available for each speaker. The dot and square markers separately show the correlation calculated based on a subset of the speech corpus or the whole corpus. The result from each large corpus is a fit by the sigmoid function.

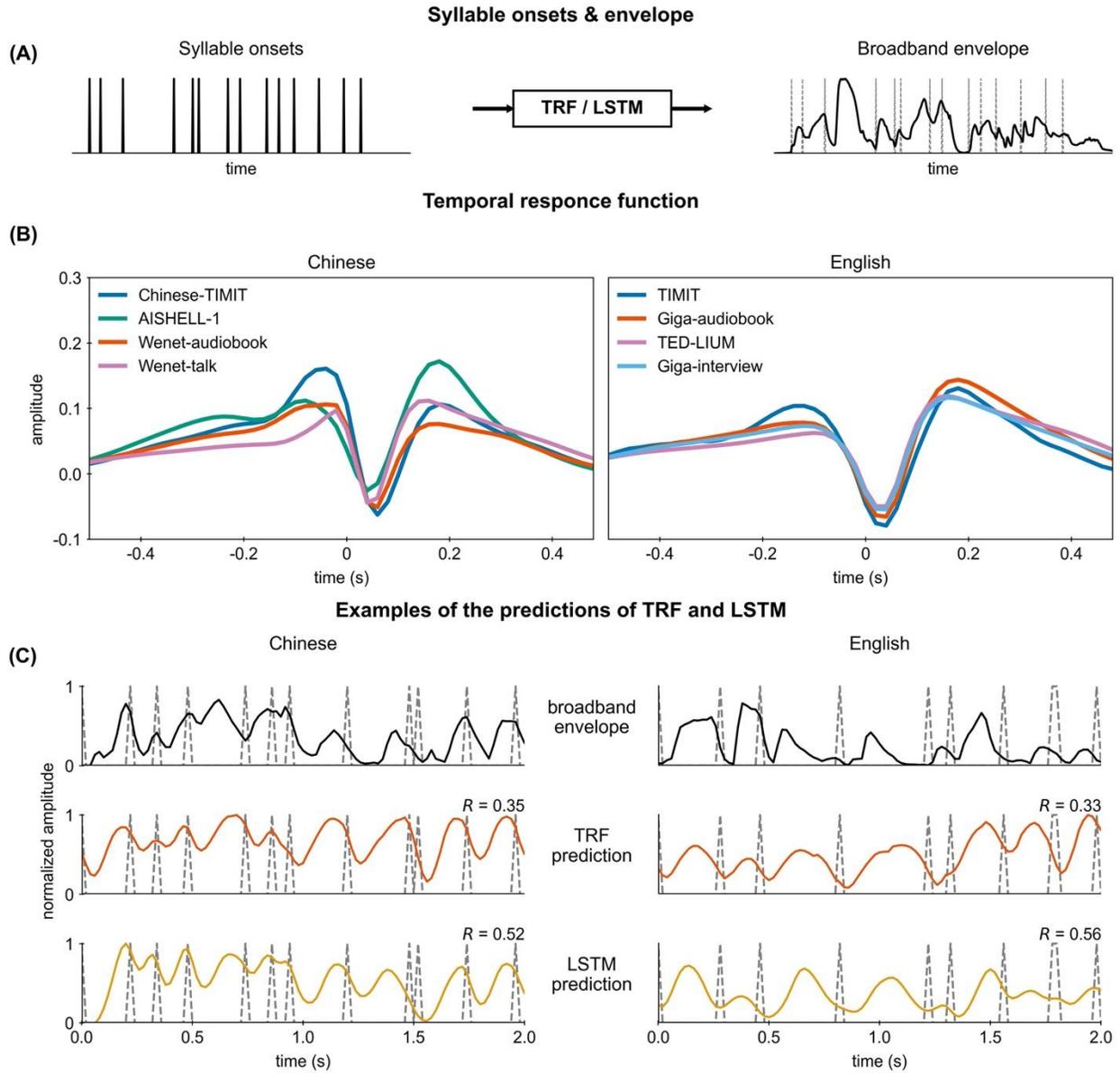

**Figure 5.** Predicting the speech envelope based on syllable onsets. (A) The syllable onset sequence is converted into the broadband envelope using either the TRF or LSTM model. (B) The TRF for each corpus is color-coded. (C) Examples of the broadband envelope predicted by the TRF and LSTM. Syllable onsets are marked by gray dashed lines. The sentences are selected so that the predictive power for the sentences is roughly consistent with the mean predictive power averaged over the whole corpus.

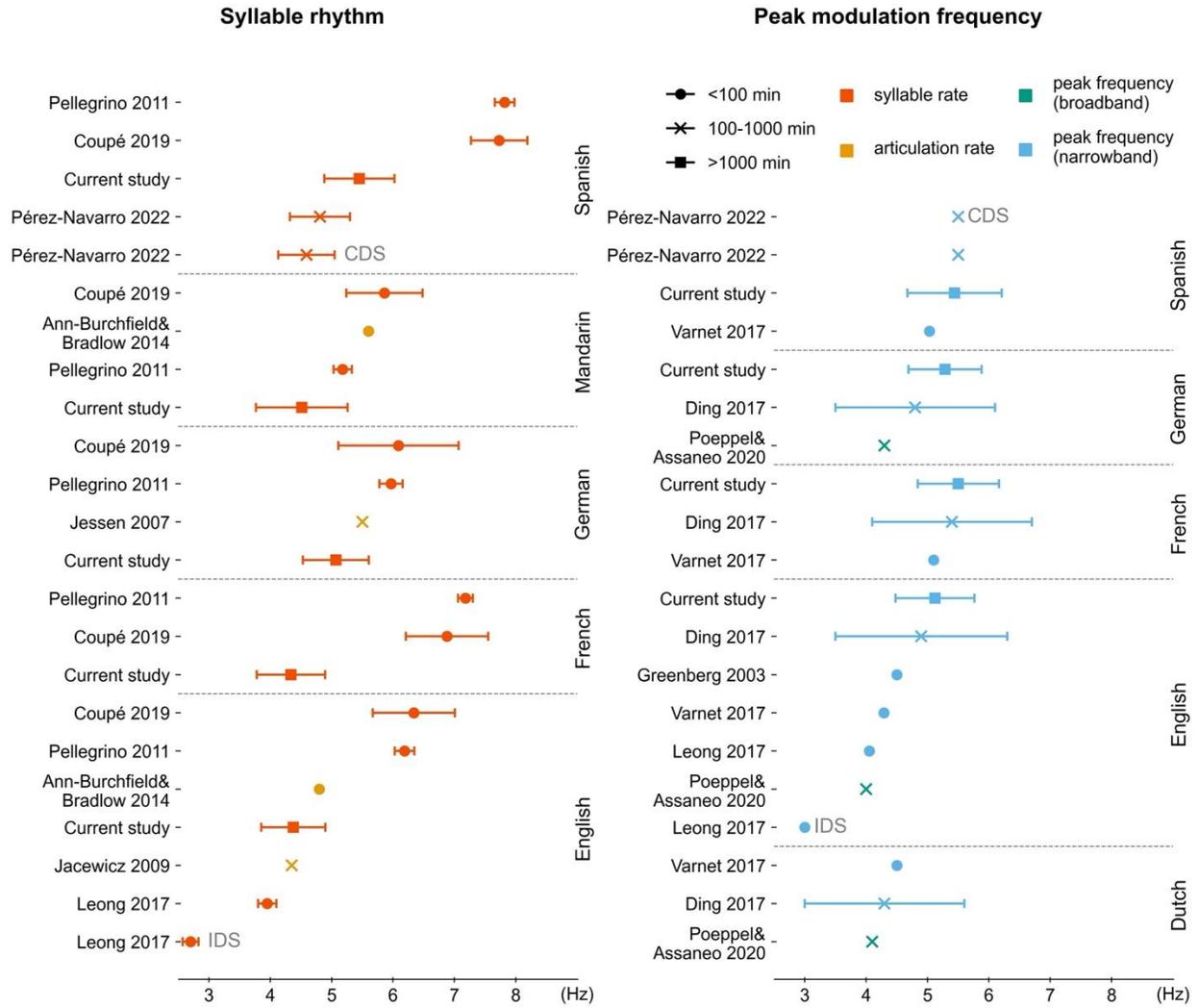

**Figure 6.** Comparison of the rate of syllables and peak modulation frequencies across studies.

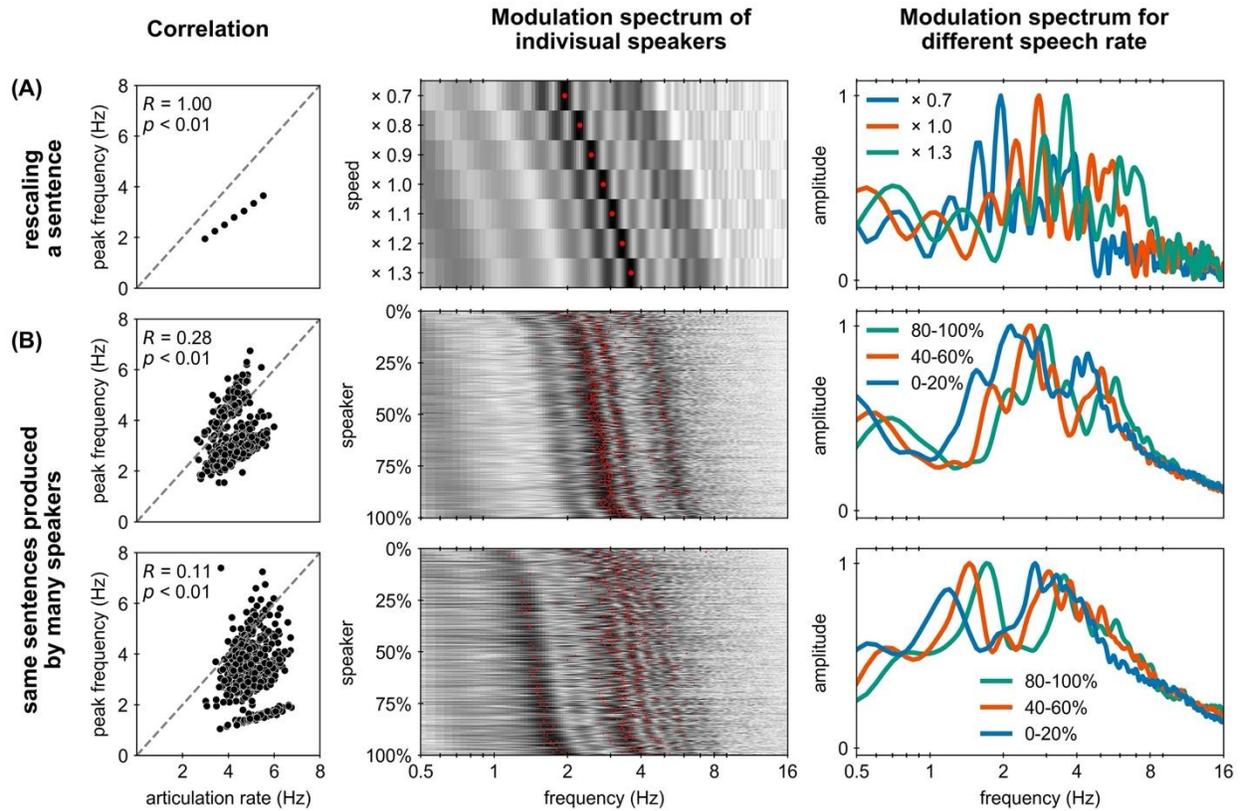

**Figure 7.** Relationship between articulation rate and modulated spectrum for example sentences. Panel A analyzes a single sentence produced by a single speaker in TIMIT (sentence ID: SA1, speaker ID: MJAC0). The sentence is artificially time compressed or expanded using the PSOLA algorithm (Boersma and Weenink, 2016). Panel B analyzes the two TIMIT sentences that are produced by many speakers (sentence ID: SA1 and SA2). The left panels show the correlation between articulation rate and peak modulation frequency. The middle panels stack the broadband modulation spectrum of each speaker, and the speakers are sorted based on the articulation rate of the sentence. The peak modulation frequency for each speaker is marked by a red dot. The right panels show representative broadband modulation spectra (A) or the spectrum averaged across different groups of speakers (B).

## Tables

### Table 1

Speech corpora.

| Name | Source | Language | Speaking Style | # sentence | Sentence Duration (s, M ± SD) | # speaker | Speaker Duration (s, M ± SD) | Total Duration (h) |
|---|---|---|---|---|---|---|---|---|
| Chinese-TIMIT | Chinese-TIMIT | Chinese | Read sentences | 6 k | 3.5 ± 0.6 | 50 | 415.1 ± 32.9 | 5.8 |
| AISHELL-1 | AISHELL-1 | Chinese | Read sentences | 142 k | 4.6 ± 1.4 | 400 | 1610.6 ± 181.2 | 179 |
| Wenet-audiobook | WenetSpeech | Chinese | Read audiobooks | 27 k | 3.2 ± 2 | 135 | 636.6 ± 201.5 | 23.9 |
| Wenet-talk | WenetSpeech | Chinese | Talk | 81 k | 2.4 ± 1.1 | 238 | 820.7 ± 254.9 | 54.3 |
| TIMIT | DARPA-TIMIT | English | Read sentences | 1 k | 2.9 ± 0.8 | 178 | 19.1 ± 2.5 | 0.9 |
| Giga-audiobook | GigaSpeech | English | Read audiobooks | 159 k | 3.8 ± 2.3 | 264 | 2303.8 ± 931.6 | 168.9 |
| TED-LIUM | TED-LIUM | English | Talk | 99 k | 6.3 ± 2.9 | 646 | 963.6 ± 157.7 | 172.9 |
| Giga-interview | GigaSpeech | English | Interviews | 31 k | 4.7 ± 3.4 | 121 | 1214.6 ± 736.2 | 40.8 |
| CV-Thai | Common-voice | Thai | Read sentences | 11 k | 4.9 ± 1.7 | 38 | 1434.5 ± 3024.1 | 15.1 |
| CV-French | Common-voice | French | Read sentences | 76 k | 4.8 ± 1.5 | 307 | 1188.8 ± 442.2 | 101.4 |
| CV-Spanish | Common-voice | Spanish | Read sentences | 32 k | 5.2 ± 1.5 | 174 | 961.1 ± 466.1 | 46.5 |
| CV-Polish | Common-voice | Polish | Read sentences | 70 k | 4.3 ± 1.5 | 190 | 1593.2 ± 3100.6 | 84.1 |
| CV-Russian | Common-voice | Russian | Read sentences | 77 k | 5.3 ± 1.7 | 226 | 1809.9 ± 4012.6 | 113.6 |
| CV-German | Common-voice | German | Read sentences | 54 k | 5.1 ± 1.6 | 400 | 693.9 ± 146 | 77.1 |

**Table 2**

Predictive power of TRF or LSTM (M ± SD).

| corpus for evaluation | TRF | | | LSTM | | |
|---|---|---|---|---|---|---|
| | trained on itself | trained on TIMIT | trained on Chinese-TIMIT | trained on itself | trained on TIMIT | trained on Chinese-TIMIT |
| Chinese-TIMIT | 0.39 ± 0.04 | 0.29 ± 0.05 | 0.39 ± 0.04 | 0.54 ± 0.03 | 0.37 ± 0.05 | 0.54 ± 0.03 |
| AISHELL-1 | 0.39 ± 0.04 | 0.37 ± 0.04 | 0.36 ± 0.04 | 0.54 ± 0.04 | 0.42 ± 0.04 | 0.47 ± 0.04 |
| Wenet-audiobook | 0.32 ± 0.05 | 0.27 ± 0.03 | 0.32 ± 0.05 | 0.54 ± 0.05 | 0.27 ± 0.06 | 0.44 ± 0.05 |
| Wenet-talk | 0.3 ± 0.05 | 0.25 ± 0.05 | 0.28 ± 0.05 | 0.45 ± 0.05 | 0.29 ± 0.04 | 0.39 ± 0.05 |
| TIMIT | 0.36 ± 0.07 | 0.36 ± 0.07 | 0.27 ± 0.06 | 0.58 ± 0.07 | 0.58 ± 0.07 | 0.33 ± 0.07 |
| Giga-audiobook | 0.32 ± 0.04 | 0.33 ± 0.04 | 0.22 ± 0.04 | 0.48 ± 0.04 | 0.43 ± 0.04 | 0.27 ± 0.06 |
| TED-LIUM | 0.28 ± 0.05 | 0.28 ± 0.05 | 0.19 ± 0.05 | 0.44 ± 0.05 | 0.3 ± 0.04 | 0.21 ± 0.05 |
| Giga-interview | 0.24 ± 0.05 | 0.25 ± 0.04 | 0.17 ± 0.04 | 0.43 ± 0.04 | 0.33 ± 0.04 | 0.22 ± 0.04 |

**Table 3**

Measures describing the rate of syllables or modulation rate (M ± SD, Hz).

| Corpus | syllable mode | syllable rate | articulation rate | broad 1 | broad 2 | narrow 1 | narrow 2 |
|---|---|---|---|---|---|---|---|
| Chinese-TIMT | 4.64 ± 0.39 | 4.96 ± 0.45 | 5.12 ± 0.45 | 4.30 ± 0.53 | 4.32 ± 0.83 | 5.25 ± 0.72 | 4.71 ± 0.75 |
| AISHELL-1 | 4.18 ± 0.47 | 4.04 ± 0.47 | 4.25 ± 0.44 | 3.89 ± 0.41 | 4.21 ± 0.63 | 4.66 ± 0.61 | 4.18 ± 0.73 |
| Wenet-audiobook | 4.43 ± 0.37 | 4.50 ± 0.46 | 5.22 ± 0.39 | 4.28 ± 0.33 | 4.12 ± 0.84 | 5.70 ± 0.49 | 5.07 ± 0.98 |
| Wenet-talk | 5.27 ± 0.64 | 5.23 ± 0.68 | 6.04 ± 0.68 | 4.87 ± 0.54 | 4.77 ± 1.14 | 6.29 ± 0.73 | 5.85 ± 1.20 |
| TIMIT | 4.31 ± 0.54 | 4.67 ± 0.51 | 4.89 ± 0.50 | 3.61 ± 0.55 | 3.37 ± 0.77 | 4.57 ± 0.78 | 4.46 ± 1.26 |
| Giga-audiobook | 3.51 ± 0.54 | 3.99 ± 0.48 | 4.30 ± 0.49 | 3.88 ± 0.44 | 3.30 ± 0.59 | 5.16 ± 0.56 | 4.65 ± 0.80 |
| TED-LIUM | 4.26 ± 0.55 | 4.44 ± 0.48 | 5.16 ± 0.52 | 4.19 ± 0.53 | 3.86 ± 0.75 | 5.24 ± 0.56 | 4.99 ± 0.88 |
| Giga-interview | 4.10 ± 0.43 | 4.44 ± 0.42 | 4.95 ± 0.42 | 3.92 ± 0.47 | 3.72 ± 0.80 | 5.25 ± 0.54 | 4.88 ± 0.98 |
| CV-Thai | / | 3.53 ± 0.29 | 3.57 ± 0.29 | 4.32 ± 0.52 | 4.38 ± 0.59 | 4.93 ± 0.60 | 4.38 ± 0.83 |
| CV-French | / | 4.34 ± 0.56 | 5.54 ± 0.60 | 5.06 ± 0.55 | 4.96 ± 0.96 | 5.50 ± 0.66 | 5.26 ± 0.95 |
| CV-Spanish | / | 5.45 ± 0.57 | 5.59 ± 0.57 | 4.88 ± 0.61 | 4.70 ± 0.91 | 5.44 ± 0.77 | 5.07 ± 0.99 |
| CV-Polish | / | 5.21 ± 0.51 | 5.39 ± 0.49 | 4.96 ± 0.56 | 4.85 ± 0.96 | 5.67 ± 0.69 | 5.35 ± 1.01 |
| CV-Russian | / | 6.14 ± 0.60 | 6.25 ± 0.59 | 4.87 ± 0.49 | 4.89 ± 0.81 | 5.40 ± 0.63 | 5.14 ± 1.04 |
| CV-German | / | 5.07 ± 0.54 | 5.12 ± 0.54 | 4.73 ± 0.55 | 4.32 ± 0.90 | 5.29 ± 0.60 | 5.00 ± 0.94 |